\documentclass[usenatbib,referee]{mn2e}
\usepackage{graphicx}
\usepackage{epsfig}
\usepackage{color}
\usepackage{enumerate}

\title[The DD model for the progenitors of SNe Ia]
{The double-degenerate model for the progenitors of type Ia supernovae}
\author[D. Liu et al.]
{D. Liu$^{\rm 1,2,3}$\thanks{E-mail:liudongdong@ynao.ac.cn}, B. Wang$^{\rm 1,2,3}$\thanks{E-mail:wangbo@ynao.ac.cn}, and Z. Han$^{\rm 1,2,3}$\thanks{E-mail:zhanwenhan@ynao.ac.cn} \\
$^1$Key Laboratory for the Structure and Evolution of Celestial Objects, Yunnan Observatories, CAS, Kunming 650216, China\\
$^2$University of Chinese Academy of Sciences, Beijing 100049, China\\
$^3$Center for Astronomical Mega-Science, CAS, Beijing 100012, China}

\begin{document}
\date{}
\pagerange{\pageref{firstpage}--\pageref{lastpage}} \pubyear{2017}
\maketitle

\label{firstpage}

\begin{abstract}\label{0. abstract}
  The double-degenerate (DD) model, involving the merging of massive double carbon-oxygen white dwarfs (CO WDs) driven by gravitational wave radiation, is one of the classical pathways for the formation of type Ia supernovae (SNe Ia). Recently, it has been proposed that the WD$+$He subgiant channel has a significant contribution to the production of massive double WDs, in which the primary WD accumulates mass by accreting He-rich matter from a He subgiant. We evolved about 1800 CO WD+He star systems and obtained a large and dense grid for producing SNe Ia through the DD model. We then performed a series of binary population synthesis simulations for the DD model, in which the WD$+$He subgiant channel is calculated by interpolations in this grid. According to our standard model, the Galactic birthrate of SNe Ia is about $2.4\times10^{\rm -3}\,\rm yr^{\rm -1}$ for the WD$+$He subgiant channel of the DD model; the total birthrate is about $3.7\times10^{\rm -3}\,\rm yr^{\rm -1}$ for all channels, reproducing that of observations. Previous theoretical models still have deficit with the observed SNe Ia with delay times $<$$1\,\rm Gyr$ and $>$$8\,\rm Gyr$. After considering the WD$+$He subgiant channel, we found that the delay time distributions is comparable with the observed results. Additionally, some recent studies proposed that the violent WD mergers are more likely to produce SNe Ia based on the DD model. We estimated that the violent mergers through the DD model may only contribute to about 16\% of all SNe Ia.
\end{abstract}

\begin{keywords}
binaries: close -- stars: evolution -- supernovae: general -- white dwarfs
\end{keywords}

\section{Introduction} \label{1. Introduction}
Type Ia supernovae (SNe Ia) are extremely powerful phenomena in the Universe, and play an exceptional role in probing the mystery of the Universe, especially for revealing the accelerating expansion of the Universe (e.g. Howell 2011; Meng, Gao \& Han 2015). The luminosities of SNe Ia are powered by the decay of $^{\rm 56}$Ni produced by the thermonuclear explosion of massive carbon-oxygen white dwarfs (CO WDs) in binaries (Hoyle \& Fowler 1960). Nevertheless, the companion stars of the exploding WDs are still elusive (Hillebrandt \& Niemeyer 2000; R\"{o}pke \& Hillebrandt 2005; Podsiadlowski et al. 2008; Wang \&Han 2012; H\"oflich et al. 2013). In the past four decades, there are two major theories dominating the landscape of SN Ia progenitor scenarios, in which the companion star is a non-degenerate star in the single-degenerate (SD) model (e.g. Whelan \& Iben 1973; Nomoto, Thielemann \& Yokoi 1984; Li \& van den Heuvel 1997; Langer et al. 2000; Han \& Podsiadlowski 2004; Chen \& Li 2007; L\"u et al. 2009) or another WD in the double-degenerate (DD) model (e.g. Iben \& Tutukov 1984; Webbink 1984; Nelemans et al. 2001). In addition, the sub-Chandrasekhar double-detonation model has been proposed to explain the sub-luminous SNe Ia, in which the companion star would be a He star or a He WD (e.g. Nomoto 1982; Woosley, Taam \& Weaver 1986).

It has been suggested that the WD$+$He star systems may produce SNe Ia via the following pathways: (1) In the SD model, the primary WD accretes He-rich matter from a He star. The accreted matter will be burned into C and O, leading to the mass increase of the primary WD. When the WD mass approaches the Chandrasekhar limit, an SN Ia is expected to be produced (Yoon \& Langer 2003; Wang et al. 2009a,b; Liu et al. 2010; Wang, Podsiadlowski \& Han 2017). In the observations, HD 49798 with its compact companion and V445 Puppis are WD$+$He star systems and may evolve to form SNe Ia in their future evolutions via the SD model (Ashok \& Banerjee 2003; Kato et al. 2008; Woudt et al. 2009; Mereghetti et al. 2009, 2011; Wang \& Han 2010; Liu et al. 2015). (2) In the DD model, a WD$+$He star system would also produce a DD system called the WD$+$He subgiant channel (see Ruiter et al. 2013; Liu et al. 2016), in which the primary WD accumulates mass via accreting He-rich matter from a He subgiant before the binary eventually evolves to a double WD system. KPD 1930+2752 may be a good candidate for producing SNe Ia through this model (Maxted, Marsh \& North 2000; Geier et al. 2007). (3) In the double-detonation model, the primary WD accretes He-rich matter at a relatively low rate, and the He-rich matter will not burn into C and O immediately but form a He-shell. A double-detonation may occur if the He-shell can be accumulated thick enough, leading to the formation of sub-luminous SNe Ia (e.g. Iben \& Tutukov 1989; Livne 1990; H\"oflich \& Khokhlov 1996; Neunteufel, Yoon \& Langer 2016). CD-30$^{\circ}$\,11223 is a WD$+$He star system that may evolve to form an SN Ia through this model (Geier et al. 2013; Wang, Justham \& Han 2013).

Recently, some observational and theoretical studies slightly favor the DD model (e.g. Howell et al. 2006; Yoon Podsiadlowski \& Rosswog 2007; Hichen et al. 2007; Scalzo et al. 2010; Chen et al. 2012; Liu et al. 2012; Horesh et al. 2012; Graham et al. 2015). Especially, the DD model have advantage of explaining the Galactic birthrates and delay time distributions of SNe Ia (e.g. Han 1998; Nelemans et al. 2001; Ruiter, Belczynski \& Fryer 2009; Maoz, Mannucci \& Nelemans 2014; Yungelson \& Kuranov 2017); the delay times of SNe Ia here are defined as the time interval from the star formation to the thermonuclear explosion. However, the delay time distributions predicted by previous theoretical studies still have deficit with the observed SNe Ia at the early epochs of $<1\,\rm Gyr$ and old epochs of $>8\,\rm Gyr$ (e.g. Yungelson \& Kuranov 2017).

According to the DD model, SNe Ia originate from the merging of double WDs with total masses larger than the Chandrasekar limit.
Note that the WD$+$He subgiant channel has been proposed as a dominant pathway for the formation of massive DD systems (e.g. Ruiter et al. 2013; Liu et al. 2016). In the present work, we added the WD$+$He subgiant channel into the DD model to explore the relative contribution of the DD model to the formation of SNe Ia. We found that the delay time distributions of SNe Ia will match better with observations after considering the WD$+$He subgiant channel, especially for SNe Ia in the old and early epochs, in which the delay time distributions were predicted to be deficient to compare with observations.

This paper will be organized as follows. In Sect.\,2, we present the numerical methods of binary evolution calculations and corresponding results. The binary population synthesis (BPS) methods and results are shown in Sect.\,3. We provided a discussion and summary in Sect.\,4.

\section{binary evolution simulations}\label{binary evolution calculations}
In order to determine whether the WD$+$He star systems evolve to form double WDs that can produce SNe Ia, we performed detailed binary evolution simulations and provided a grid for producing SNe Ia via the DD model. This grid can be constructed to make interpolations in the subsequent BPS studies (see Sect.\,3).

\subsection{Numerical methods}
Employing Eggleton's stellar evolution code that has undergone many revisions (Eggleton 1973; Han, Podsiadlowski \& Eggleton 1994; Pols et al. 1995, 1998; Eggleton \& Kiseleva-Eggleton 2002), we evolved a large number of WD$+$He star systems. The ratio of the mixing length to the local pressure scale height (i.e. $\alpha$$=$$l/H_{\rm p}$) is assumed to be 2.0. The initial He star models are assumed to have a He mass fraction $Y$$=$$0.98$ and metallicity $Z$$=$$0.02$. The loss of orbital angular momentum caused by the gravitational wave radiation is also included (see Landau \& Lifshitz 1971).

When the He star evolves to fill their Roche lobe, the mass transfer process begins.
The process of Roche lobe overflow (RLOF) is treated similar to that described in Han, Tout \& Eggleton (2000), in which the mass donor star is assumed to overflow its Roche lobe stably and only when needed, but never too much. According to the prescriptions in Nomoto (1982) and Kato \& Hachisu (2004), we calculated the mass growth rate of the WD ($\dot{M}_{\rm WD}$) as described below:
\begin{equation}
\dot{M}_{\rm WD}=\left\{
\begin{array}{lcl}
\dot{M}_{\rm cr},\,\,\,\,\,\,\,\,\,\,\,\,\,\,\,\, \dot{M}_{\rm 2}>\dot{M}_{\rm cr}\\
\dot{M}_{\rm 2},\,\,\,\,\,\,\,\, \dot{M}_{\rm st}<\dot{M}_{\rm 2}<\dot{M}_{\rm cr}\\
\eta_{\rm He}\dot{M}_{\rm 2},\,\,\,\,\,\,\,\, \dot{M}_{\rm 2}<\dot{M}_{\rm st}
\end{array}
\right.
\end{equation}
where $\dot{M}_{\rm cr}$ is the critical mass transfer rate (see Nomoto 1982), $\dot{M}_{\rm 2}$ is the mass transfer rate, $\dot{M}_{\rm st}$ is the minimum accretion rate for stable He shell burning and $\eta_{\rm He}$ is the mass accumulation efficiency for He shell flash (see Kato \& Hachisu 2004). In the case of $\dot{M}_{\rm 2}>\dot{M}_{\rm cr}$, the transferred He-rich matter is assumed to burn into C and O stably at the rate of $\dot{M}_{\rm cr}$, whereas the rest of the He-rich matter is assumed to be blown away from the system in the form of optically thick wind at the rate of $(\dot{M}_{\rm 2}-\dot{M}_{\rm cr})$ (see Hachisu, Kato \& Nomoto 1996).

After the formation of double WD systems, the double WDs are gradually approaching and eventually merge due to the gravitational wave radiation.
In the present work, we assume that all mergers of double CO WD systems with total masses larger than the Chandrasekhar limit (set to be $1.378\,\rm M_{\odot}$) would form SNe Ia, aiming to estimate the possible contribution of the DD model to the birthrates and delay time distributions of SNe Ia. Furthermore, the delay times of SNe Ia are constrained to be shorter than the Hubble time. Note that the criteria for the WD mergers to producing SNe Ia is still under debate (e.g. Dan et al. 2012, 2014; Sato et al. 2016; Yungelson \& Kuranov 2017). We will also discussed the influence of different critical mass ratios of WD mergers on the final results.

We incorporate the prescriptions above into Eggleton's stellar evolution
code to simulate the evolution of WD$+$He star systems. We assume that the mass loss
from these systems will take away specific orbital
angular momentum of the accreting WDs. We have simulated the
evolution of about 1800 WD$+$He star systems, and thereby obtained a dense
and large model grid for producing SNe Ia via the DD model. This grid can be used in BPS simulations and for searching SN Ia progenitor candidates. In our calculations, the
range of initial masses of the primary WDs ($M_{\rm WD}^{\rm i}$) is from $0.5$ to $1.2\,\rm M_{\odot}$, which is almost the maximum mass range for CO WDs; the initial masses of the He stars ($M_{\rm 2}^{\rm i}$)
varies from $0.3$ to $2.3\,\rm M_{\odot}$; for the initial orbital
periods of the binaries ($P^{\rm i}$), the He stars fill their Roche lobe at the He zero-age main sequence (MS) stage beyond the left boundary, whereas the delay times for the formation of double WD mergers is larger than the Hubble time beyond the right boundary.

\subsection{A typical binary evolution example}
In Fig.\,1, we present a representative example for the
binary evolution of a WD$+$He star system that will evolve to a massive DD system and eventually form an SN Ia. In this figure, the left panel presents the mass transfer rate from the He star
($\dot{M}_{\rm 2}$), the mass growth rate of the WD ($\dot{M}_{\rm WD}$) and the mass of the WD ($M_{\rm WD}$) as a function of
time, and the right panel shows the evolutionary track of the He
star in the Hertzsprung-Russell diagram and the evolution of the
orbital periods. The initial parameters of this WD$+$He star system
are ($M_{\rm WD}^{\rm i}$, $M_{\rm 2}^{\rm i}$, $\log\,P^{\rm i})=(0.9,
1.2, -0.7)$, in which $M_{\rm WD}^{\rm i}$ and $M_{\rm He}^{\rm i}$ are in units of $\rm M_{\odot}$, and $P^{\rm i}$ is in units of days.

\begin{figure*}
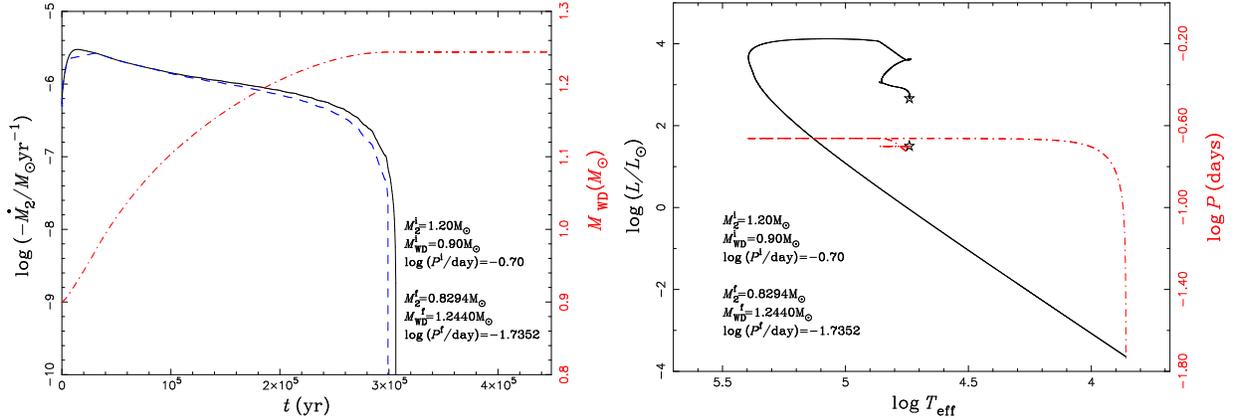

\centerline{\epsfig{file=f1a.ps,angle=270,width=8cm}\ \ \epsfig{file=f1b.ps,angle=270,width=8cm}} \caption{The
  evolution of a representative WD$+$He star system for
  producing double CO WD system that potentially form an SN Ia. The left panel shows
  the evolution of the mass transfer rate (solid line), the WD
  mass-growth rate (dashed line) and the WD mass (dash-dotted line)
  changing with time. In the right panel, the solid line and the dash-dotted line show the
  luminosity of the mass donor and the binary orbital
  period as a function of effective temperature, respectively. Asterisks in the right panel indicate the position where the simulation begins.}
\end{figure*}

We start our simulation when the companion of the WD is at the He MS stage. After about $1.0 \times 10^7\,\rm yr$, the He star fills its Roche lobe at the subgiant stage, resulting in a mass transfer process from the He star onto the surface of WD. At the beginning, the mass transfer rate is about $2.5 \times 10^{\rm-7}\,\rm M_{\odot}yr^{\rm -1}$ and keep increasing. After about $6 \times 10^3\,\rm yr$, the mass transfer exceeds the critical rate $\dot{M}_{\rm cr}$ and enters a stellar wind stage. During this phase, the transferred He-rich matter burns into C and O and accumulates onto the surface of the WD at the rate of $\dot{M}_{\rm cr}$, whereas the rest of the He-rich matter is blown away from the system in the form of optically thick wind at the rate of $(\dot{M}_{\rm 2}-\dot{M}_{\rm cr})$. The mass transfer rate reduces to be lower than $\dot{M}_{\rm cr}$ and the stellar wind stops about $2.5 \times 10^4\,\rm yr$ later. During this phase, the transferred He-rich matter burn stably and no matter is lost from the system. After about $5.5 \times 10^4\,\rm yr$, the binary system evolves into a weak He-shell flash phase and the WD mass still increases. When the envelope of the He star is exhausted about $2.0 \times 10^5\,\rm yr$ later, the He star turns to be a WD and thus a double WD system is formed. Finally, the mass of the primary WD is $M_{\rm WD}^{\rm f}=1.2440\,\rm M_{\odot}$, the mass of the WD originating from the He star is $M_{\rm 2}^{\rm f}=0.8294\,\rm M_{\odot}$, and the orbital period is $\log(P^{\rm f}/\rm day)=-1.7352$ about $2.0 \times 10^9\,\rm yr$ later. Subsequently, the double WDs continues to cool and the system will eventually merge in about $11\,\rm Myr$ driven by the gravitational wave radiation, leading to the formation of an SN Ia explosion.

\subsection{Initial parameters for SNe Ia}
\begin{figure*}
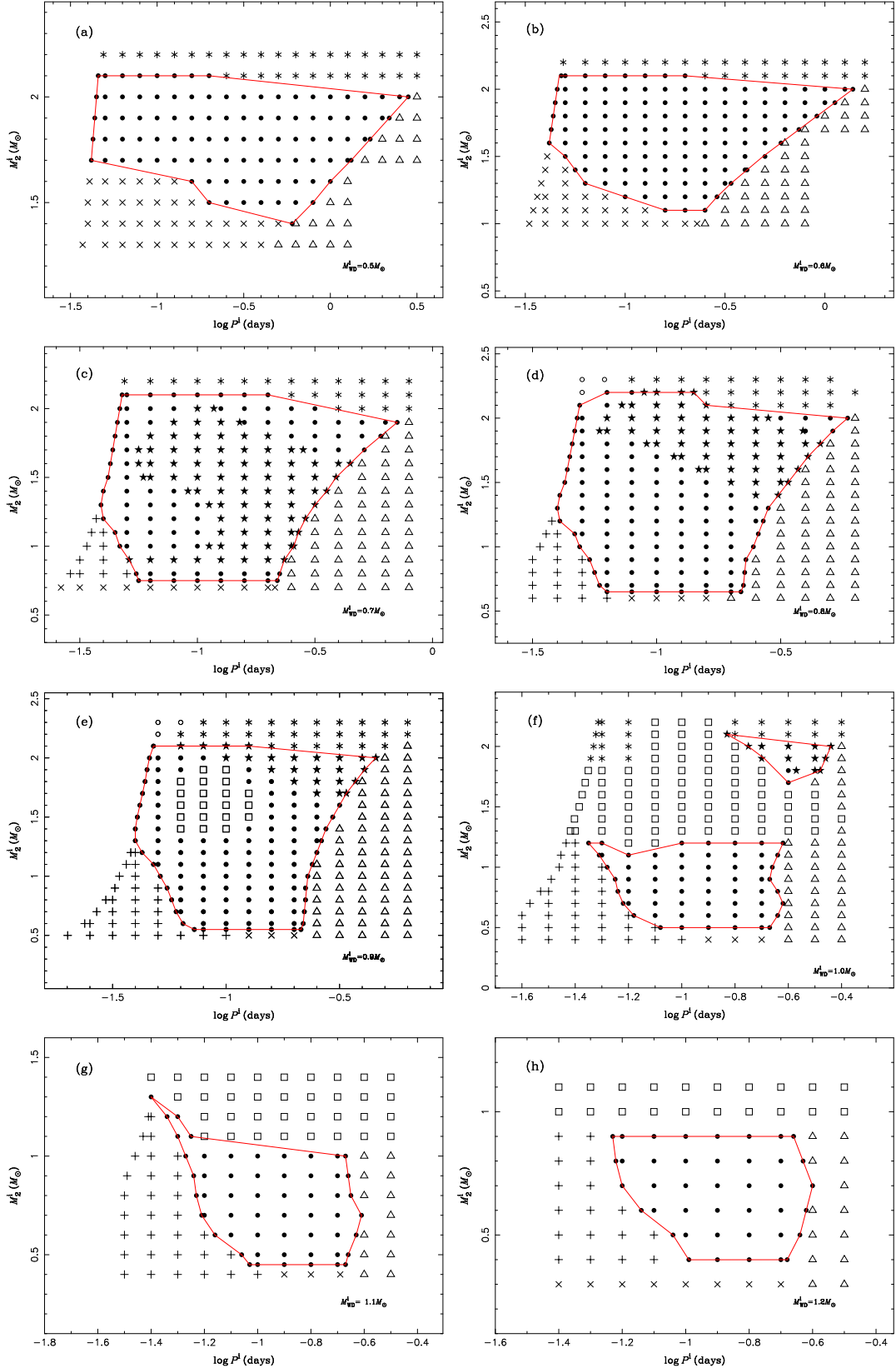

\begin{tabular}{@{}cc@{}}
\centerline{\epsfig{file=f2a.ps,angle=270,width=7.cm}\ \ \epsfig{file=f2b.ps,angle=270,width=7.cm}} \\
\centerline{\epsfig{file=f2c.ps,angle=270,width=7.cm}\ \ \epsfig{file=f2d.ps,angle=270,width=7.cm}} \\ \centerline{\epsfig{file=f2e.ps,angle=270,width=7.cm}\ \ \epsfig{file=f2f.ps,angle=270,width=7.cm}} \\ \centerline{\epsfig{file=f2g.ps,angle=270,width=7.cm}\ \ \epsfig{file=f2h.ps,angle=270,width=7.cm}}
\end{tabular}
\caption{The grid of WD$+$He star systems for producing SNe Ia in the $\log P^{\rm i}-M^{\rm i}_2$ plane. Every panel shows the outcomes of a particular initial WD mass. The filled circles and filled five-pointed stars in red contours represent binaries that would produce SNe Ia via the classical DD model and violent merger scenario, respectively. Other binaries cannot evolve to form SNe Ia via the DD model; they may form double WDs less massive than the Chandrasekhar limit (crosses), or form ONe WD$+$CO WD systems (asterisks), or have delay times larger than the Hubble time (triangles), or produce SNe Ia via the double-detonation model (pluses) or the SD model (squares), or undergo a dynamical unstable mass transfer process (open circles).}
\end{figure*}

Having evolved a large number of WD$+$He star systems until the onset of the merging of double WDs, we provided the final outcomes of WD$+$He star systems in the initial orbital period$-$initial He star mass ($\log P^{\rm i}-M^{\rm i}_2$) plane in Fig.\,2. In order to produce SNe Ia via the DD model, the masses of CO WDs are from $0.5$ to $1.2\,\rm M_{\odot}$, the masses of He stars range from $0.4$ to $2.2\,\rm M_{\odot}$ and the orbital periods change from $0.96\,\rm h$ to $2.82\,\rm day$. In this figure, the filled circles represent WD$+$He star systems that can produce SNe Ia in their future evolution based on the classical DD model, while the filled five-pointed stars indicate that these WD$+$He star systems will evolve to double WDs potentially merge violently and produce SNe Ia (see Liu et al. 2016). We also found that the primary WDs can obtain mass up to $0.48\,\rm M_{\odot}$ by accreting He-rich matter from the He donors.

In Fig.\,2, the WD$+$He star systems that locate outside of the contours cannot produce SNe Ia through the DD mode, as follows:

\begin{enumerate}[(1)]
  \item Binary systems denoted by crosses: double WDs produced from these systems have total masses lower than the Chandrasekhar limit. Those systems are expected to evolve to be a single massive CO WD. RE J0317$-$853, which is a magnetic WD rotating rapidly with its mass close to the Chandrasekhar limit, seems to be a good candidate from the merging of double CO WDs (Barstow et al. 1995; K\"ulebi et al. 2010; Tout et al. 2008; Maoz, Mannucci\& Nelemans 2014).

  \item Binary systems represented by asterisks: massive He stars in these binaries will evolve to be ONe WDs but not CO WDs, i.e. forming CO WD$+$ONe WD systems. The merging of these systems may collapse as a neutron star, which should be surrounded by C and O rich ejections. 

  \item Binary systems marked by triangles: the delay times for the production of mergers of the DD systems originating from these systems are larger than the Hubble time.

  \item Binary systems indicated by pluses: the He stars will fill their Roche lobe when they are at the He MS stage since the initial separations of these binaries are relatively close. In this case, the mass transfer rate is relatively low, leading to the formation of a thick layer of helium on the surface of the WD. The detonation of the thick He layer may trigger the explosion of the whole WD (i.e. the double-detonation model), corresponding to sub-luminous SNe Ia (Nomoto 1982; Woosley et al. 1986; Wang, Justham \& Han 2013). 

  \item Binary systems represented by squares: the primary WDs in these binaries will increase their mass to the Chandrasekhar limit and produce SNe Ia via the SD model (see Wang et al. 2009a).

  \item Binary systems denoted by open circles: the mass transfer is dynamically unstable when the He subgiants fill their Roche lobe, resulting in a common envelope (CE) phase. After the ejection of CE, these systems may also evolve to double WDs and produce SNe Ia based on the DD model. We will discuss this scenario as part of the CE ejection channels in the Sect.\,3.
\end{enumerate}

In Fig.\,3, we present the parameter space of WD$+$He star systems that can produce SNe Ia through the DD model in the $\log P^{\rm i}-M_{\rm 2}^{\rm i}$ plane, in which the initial WD mass $M_{\rm WD}^{\rm i}$ varies from $0.5$ to $1.2\rm\, M_{\odot}$ for different contours. From this figure, we can see that the contours turn to move to upstairs for lower initial WD masses, which is caused by the requirement that the total mass of double WDs should be larger than the Chandrasekhar limit for producing SNe Ia. In this figure, the square with error bar represents a WD$+$sdB star system KPD 1930+2752, the location of which implies that KPD 1930+2752 would produce an SN Ia via the DD model. More detailed calculations on the future evolution of KPD 1930$+$2752 see Sect.\,4.

\begin{figure}
\begin{center}
\epsfig{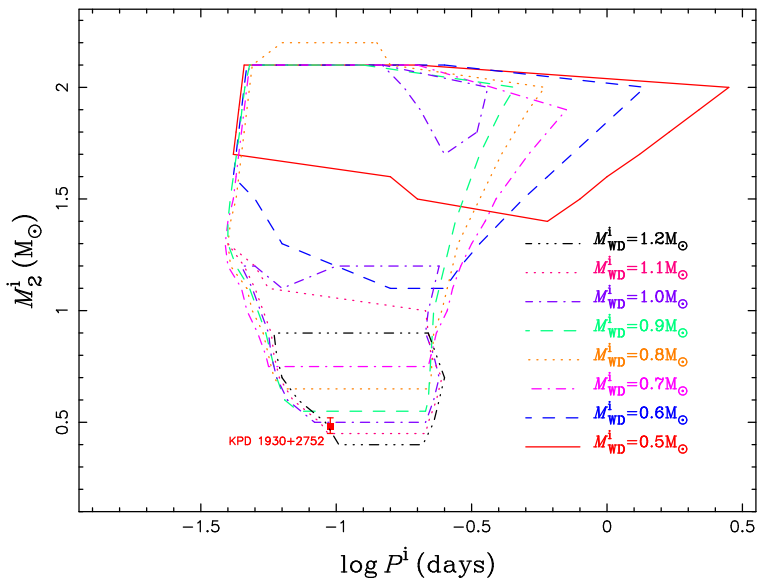}
 \caption{Parameter space of WD$+$He star systems for producing SNe Ia based on the DD model.}
  \end{center}
\end{figure}

\section{Binary population synthesis}\label{Binary population synthesis}
\subsection{BPS approach}
In order to calculate the birthrates and delay times of SNe Ia, we combined Hurley's rapid binary evolution code (Hurley, Tout \& Pols 2002) with the Monte Carlo method. The solar metallicity $Z=0.02$ is adopted in our calculations. In each of our simulation, we evolved $1\times10^7$ primordial binaries from their formation to the formation of WD$+$He star systems. If the parameters of the formed WD$+$He star systems are located in the parameter space presented in Fig.\,3. We investigated the properties of the double WDs at the moment of their formation time by interpolations in the grids from binary evolution calculations.

In this work, the formation of WD$+$He star systems are mainly through the following two channels (see also Liu et al. 2016). (1) \emph{Channel} A. When the primordial primary evolves to the subgiant stage, it fills its Roche lobe and transfers H-rich envelope stably onto the primordial secondary. After the mass transfer process, the primordial primary turns to be a He star. Subsequently, the secondary evolves to a subgiant star and fills its Roche lobe, forming a CE since the mass transfer is dynamical unstable. A He subgiant$+$He star system is produced after the CE ejection. The He subgiant quickly fills its Roche lobe again and enters a stable mass transfer process. After that, a CO WD$+$He star is formed. (2) \emph{Channel} B. Similar to \emph{Channel} A, the primordial primary first evolve to a He star and continues to evolve. When the He star evolves to the He subgiant stage, it fills its Roche lobe and transfer He-rich matter stably onto the MS star. After that, a CO WD$+$MS will be formed. When the MS star will evolve to a subgiant star, it fills its Roche lobe and the mass transfer is dynamically unstable, resulting in a CE process. If the CE can be ejected, a CO WD$+$He star system will be formed.

There are some basic assumption and initial parameters as input of the Monte Carlo BPS calculations, as follows:
(1) We assumed that all stars are members of binaries and their orbits are circular.
(2) The mass distribution of the primordial primary stars is assumed to follow the initial mass function described in Miller \& Scalo (1979).
(3) The mass ratio distribution is taken to be constant.
(4) The distribution of initial orbital separations are assumed to be divided into two parts: one is constant in $\log\,a$ for the wide binaries with orbital period lager than 100 yr, and the other one falls off smoothly for the close binaries, where $a$ is the orbital separation (see Eggleton, Fitchett \& Tout 1989; Han, Podsiadlowski \& Eggleton 1995). The number of binaries in the wide part and close part is assumed to be equal.
(5) In order to provide a approximate description for spiral galaxies and elliptical galaxies, we simply employ a constant star formation rate (SFR) and a delta-function SFR, respectively. For the situation of our Galaxy, the SFR is calibrated to be $5\,\rm M_{\odot}\,\rm yr^{\rm -1}$ over the past $15\,\rm Gyr$ (Yungelson \& Livio 1998; Willems \& Kolb 2004; Han \& Podsiadlowski 2004). A star burst of $10^{10}\,\rm M_{\odot}$ in stars is assumed for the case of elliptical galaxies.

The WD$+$He star systems in the present work generally evolve from the CE ejections. However, the prescription for CE ejection is still under debate (e.g. Ivanova et al. 2013). In order to calculate the output of the CE phase, we employed the standard energy prescription described in Webbink (1984), in which the CE ejection efficiency $\alpha_{\rm CE}$ and the stellar structure parameter $\lambda$ are two inconclusive parameters. Similar to Liu et al. (2016), we simply combined these two parameters as a single free parameter (i.e. $\alpha_{\rm CE}\lambda$) to examine its influence on the final results. Here, we set $\alpha_{\rm CE}\lambda=1.0$ in our standard model and change it to 0.5 and 1.5 for comparison.

\subsection{BPS Results}
\subsubsection{Birthrates and delay times of SNe Ia}
\begin{table*}
\begin{center}
 \caption{The primordial parameters and BPS results for different simulation sets based on the DD model. Notes: The mass ratio $q_{\rm cr}$ is the minimum $q=M_{\rm WD2}/M_{\rm WD1}$ assumed for producing SNe Ia, where $M_{\rm WD1}$ and $M_{\rm WD2}$ are the mass of the massive WD and the less massive WD, respectively. $M_{\rm 10}$, $M_{\rm 20}$ and $P_{\rm 0}$ are the initial mass of the primordial primary, the initial mass of the primordial secondary and the primordial orbital period, respectively.}
   \begin{tabular}{ccccccccc}
\hline \hline
 Set & $Channels$ & $\alpha_{\rm CE}\lambda$ & $q_{\rm cr}$ & $M_{\rm 10}$ & $M_{\rm 20}$ & $P_{\rm 0}$ & $\rm Birthrates$ & $\rm DTDs$\\
 & & & & $(\rm M_{\odot})$ & $(\rm M_{\odot})$ & (days) & $(10^{\rm -3}{\rm yr}^{\rm -3})$ & $({\rm Myr})$\\
\hline
$1$ & $\rm WD+He\,\,subgiant$ & $0.5$ & $\rm Unlimited$ & $3.0-8.5$ & $1.5-6.0$ & $5-6500$ & $0.77$ & $>140$\\
$2$ & $\rm WD+He\,\,subgiant$ & $1.0$ & $\rm Unlimited$ & $3.0-8.5$ & $1.5-6.5$ & $2-4900$ & $2.39$ & $>110$\\
$3$ & $\rm WD+He\,\,subgiant$ & $1.5$ & $\rm Unlimited$ & $3.0-9.5$ & $1.5-7.5$ & $2-4900$ & $3.70$ & $>110$\\
$4$ & $\rm CE\,\,ejection$ & $0.5$ & $\rm Unlimited$ & $3.0-6.5$ & $1.5-6.5$ & $1200-7200$ & $0.48$ & $>70$\\
$5$ & $\rm CE\,\,ejection$ & $1.0$ & $\rm Unlimited$ & $3.0-8.5$ & $1.0-6.5$ & $4-6400$ & $1.34$ & $>70$\\
$6$ & $\rm CE\,\,ejection$ & $1.5$ & $\rm Unlimited$ & $3.0-9.5$ & $1.0-7.0$ & $3-5600$ & $2.93$ & $>89$\\
$7$ & $\rm All$ & $1.0$ & $\rm Unlimited$ & $3.0-8.5$ & $1.0-6.5$ & $2-6400$ & $3.73$ & $>70$\\
$8$ & $\rm All$ & $1.0$ & $0.8$ & $3.0-8.5$ & $1.0-6.5$ & $3-4900$ & $0.65$ & $>70$\\
$9$ & $\rm All$ & $1.0$ & $0.9$ & $3.5-8.5$ & $1.5-6.0$ & $3-4900$ & $0.17$ & $>70$\\
\hline \label{1}
\end{tabular}
\end{center}
\end{table*}
In the WD$+$He subgiant channel, the mass transfer before the formation of DD system is dynamically stable, so this channel also can be named as the stable mass transfer channel. In addition, there are some other channels for the formation of massive DD systems that can produce SNe Ia, in which all DD systems originate from the CE ejection process (e.g. Han 1998; Toonen, Nelemans \& Portegies 2012; Meng \& Yang 2012; Yungelson \& Kuranov 2017). Thus, we named these other channels as the CE ejection channels in the present work. In Table\,1, we show the birthrates and delay time distributions (DTDs) of SNe Ia from the WD$+$He subgiant channel and CE ejection channels with different CE ejection parameters. As the mass ratio of WD mergers may influence their outcomes, we show the influence of the mass ratio criteria of the merging WDs on our final results of all channels here. We also present the range of initial binary parameters for producing SNe Ia in Table\,1.

In Fig.\,4, we show the Galactic birthrate evolution of SNe Ia as a function of time based on the DD model. In this figure, the thick lines represent the birthrates of SNe Ia from all channels of the DD model, while the thin lines show that from the WD$+$He subgiant channel. Here, the star formation rate is set to be constant ($5\,\rm M_{\odot}yr^{\rm -1}$) and the metallicity is assumed to be 0.02. According to our standard model ($\alpha_{\rm CE}\lambda=1.0$), the Galactic SN Ia birthrate of the WD$+$He subgiant channel is $\sim$$2.4\times10^{\rm -3}\,\rm yr^{\rm -1}$, and the total birthrate of SNe Ia from all channels of the DD model is $\sim$$3.7\times10^{\rm -3}\,\rm yr^{\rm -1}$. This is consistent with the observed results ($3-4\times10^{\rm -3}\,\rm yr^{\rm -1}$, e.g. Cappellaro \& Turatto 1997). The WD$+$He subgiant channel contributes about 64\% of all SNe Ia from the DD model based on our standard model, which verifies that the CO WD$+$He star channel is one of the most important channel for the formation of massive DD systems. We also note that the birthrate of SNe Ia from the DD model increase with the $\alpha_{\rm CE}\lambda$, which is caused by the fact that less energy will be required during the CE ejection when a larger $\alpha_{\rm CE}\lambda$ is adopted, leading to the formation of more WD$+$He star systems that can produce SNe Ia based on our assumptions.

\begin{figure}
\begin{center}
\epsfig{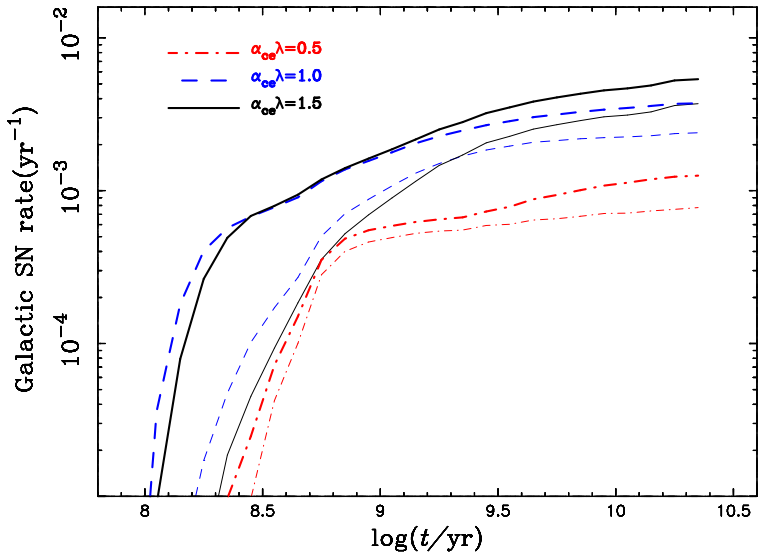}
 \caption{Evolution of Galactic SN Ia birthrates as a function of time based on the DD model. The thick lines represent the case producing SNe Ia from all channels of the DD model, while the thin lines show the case from the WD$+$He subgiant channel.}
  \end{center}
\end{figure}

Fig.\,5 presents the theoretical DTDs of SNe Ia from the merging of double WDs originating from different channels. Here, we show the results with $\alpha_{\rm CE}\lambda=1.0$ (left panel) and 1.5 (right panel). We assume a star burst of $10^{10}\,\rm M_{\odot}$ in stars here and set the metallicity $Z=0.02$.
The delay times of SNe Ia range from $110\,\rm Myr$ to the Hubble time based on the WD$+$He subgiant channel, and from $70\,\rm Myr$ to the Hubble time based on all channels. The cut-offs at the large end of DTDs are artificial since the system ages have already reached the Hubble time.
For SNe Ia from all channels, the DTDs are roughly proportional to $t^{\rm -1}$ and the rate of SNe Ia is comparable with that in the elliptical galaxies (or galaxy clusters) from observations. Similar to previous studies (e.g. Yungelson \& Kuranov 2017), the CE ejection channels still have deficit compared with observations for SNe Ia in early and old epochs. After considering the WD$+$He subgiant channel, we found that the DTDs here can match better with the observations. Note that the cases with $\alpha_{\rm CE}\lambda=0.5$ is hard to reproduce the DTDs in observations, which indicates that a larger value of $\alpha_{\rm CE}\lambda$ is needed if the DD model can contribute to most SNe Ia.

\begin{figure}
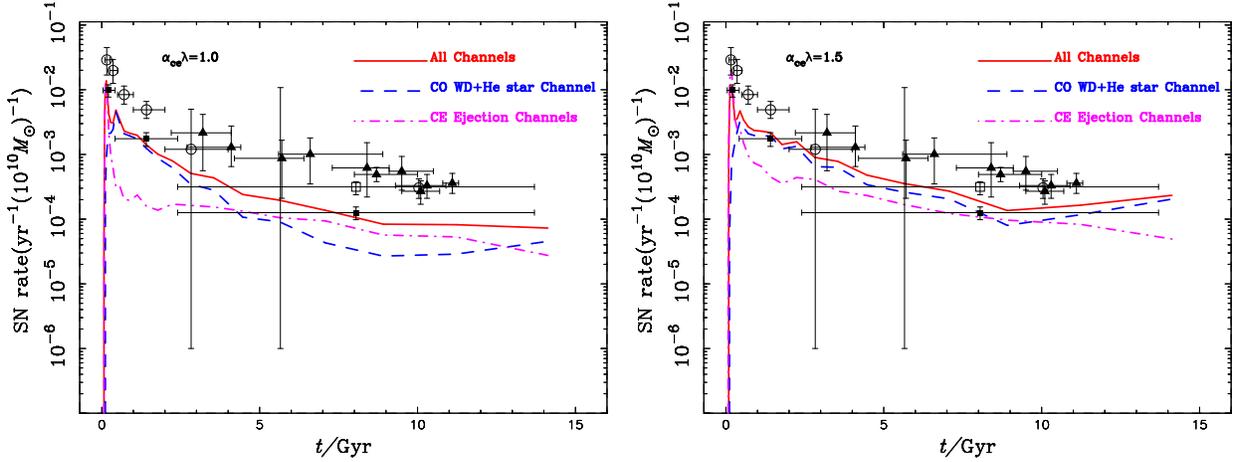

\begin{center}
\begin{tabular}{@{}c@{}}
\centerline{\epsfig{file=f5a.ps,angle=270,width=8cm} \ \ \epsfig{file=f5b.ps,angle=270,width=8cm}}
\end{tabular}
 \caption{Delay time distributions of SNe Ia from different formation channels of the DD model. The $\alpha_{\rm CE}\lambda$ is set to be 1.0 (left panel) and 1.5 (right panel). Points with error bars represent the observational date of elliptical galaxies from the Subaru/XMM$-$Newton Deep Survey (the open circles; Totani et al. 2008), galaxy clusters at the redshifts from $Z=0$ to $Z=1.45$ (the filled triangles; Maoz, Keren \& Avishay 2010), a galaxy sample from SDSS (the filled squares; Maoz, Mannucci \& Timothy 2012), and an SN Ia sample from the Cluster Lensing And Supernova survey with Hubble (the open square; Graur \& Maoz 2013).}
\end{center}
\end{figure}

\subsubsection{Mass distributions of double WDs}
In Fig.\,6, we present the density distribution for the masses of double WDs potentially producing SNe Ia in the $M_{\rm WD}^{\rm f}-M_{\rm 2}^{\rm f}$ plane, where $M_{\rm WD}^{\rm f}$ and $M_{\rm 2}^{\rm f}$ are the final mass of the primary WD and the mass of the WD originating from the He star, respectively. In most cases, $M_{\rm WD}^{\rm f}$ is larger than $M_{\rm 2}^{\rm f}$, whereas the fraction for $M_{\rm WD}^{\rm f}<M_{\rm 2}^{\rm f}$ is about 2.5\%. It is worth noting that this distribution divided into two parts: $M_{\rm 2}^{\rm f}$ is concentrated in the vicinity of $0.7\,\rm M_{\odot}$ for the primary WDs in the mass range of $0.8-1.25\,\rm M_{\odot}$, and $M_{\rm 2}^{\rm f}$ mainly distribute in the range of $0.8-1.0\,\rm M_{\odot}$ for the primary WDs with masses larger than $1.25\,\rm M_{\odot}$. This divided distribution is caused by the WD$+$He subgiant channel, whereas double WDs originating from the CE ejection channels are totally located in the less massive part. According to the WD$+$He subgiant channel, double WDs in the less massive part are mainly originated from WD$+$He star system produced via \emph{Channel}\,A, while the massive part are formed via \emph{Channel}\,B (see Sect.\,3.1). 
The filled squares with error bars in Fig.\,6 represent the DD core of a planetary nebula Henize 2-428 and the double WDs that would evolve from KPD 1930+2752, respectively.
Henize 2-428 is a planetary nebula with double degenerate core that has a total mass of $\sim$$1.76\,\rm M_{\odot}$, mass ratio $q$$\sim$$1$ and orbital period $\sim$$4.2\,\rm h$ (Santander-Garc\'ia et al. 2015), which is a good candidate for violent merger scenario of SN Ia progenitors. In addition, the merging of another super-Chandrasekhar DD system NLTT 12758 may also lead to an SN Ia based on the violent merger scenario, although the gravitational wave radiation time for its merging is larger than the Hubble timescale (Kawka et al. 2017).


\begin{figure*}
\begin{center}
\begin{tabular}{@{}c@{}}
\centerline{\epsfig{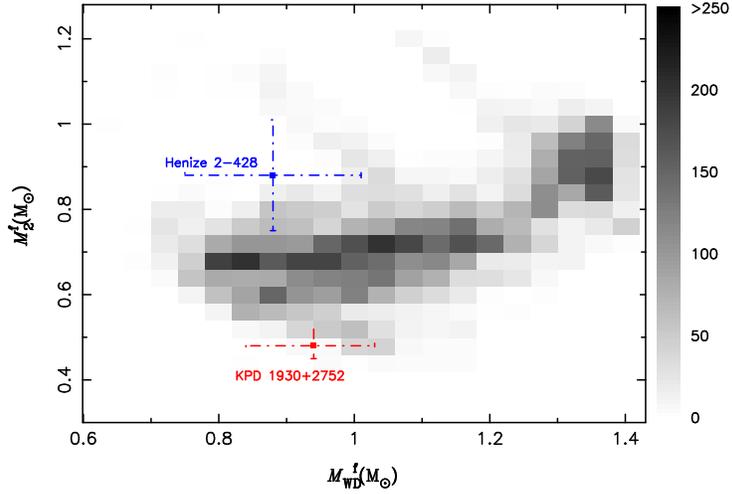}}
\end{tabular}
 \caption{The density distribution of WD mergers potentially producing SNe Ia via the DD model in the $M_{\rm WD}^{\rm f}-M_{\rm 2}^{\rm f}$ plane. Here, the $\alpha_{\rm CE}\lambda$ is set to be 1.0.}
  \end{center}
\end{figure*}

\begin{figure*}
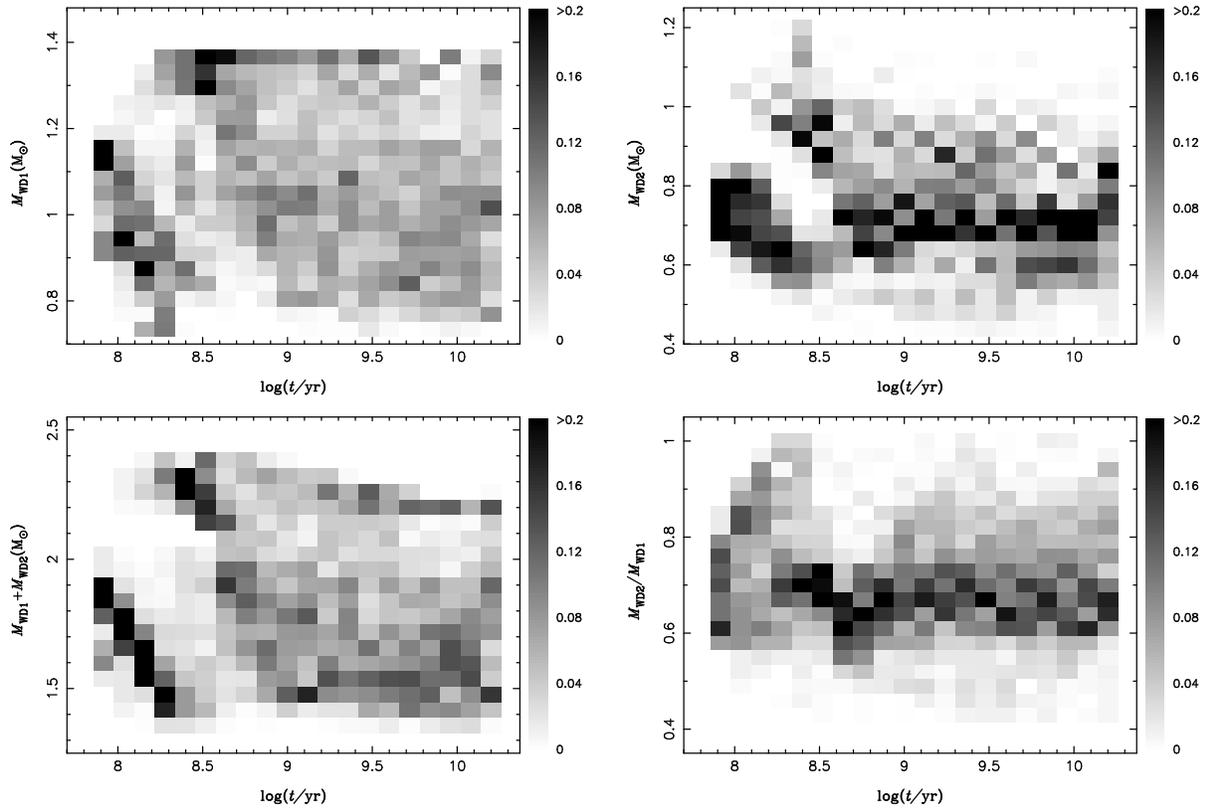

\begin{center}
\begin{tabular}{@{}cc@{}}
\centerline{\epsfig{file=f7a.ps,angle=270,width=8cm}\ \ \epsfig{file=f7b.ps,angle=270,width=8cm}} \\
\centerline{\epsfig{file=f7c.ps,angle=270,width=8cm}\ \ \epsfig{file=f7d.ps,angle=270,width=8cm}}
\end{tabular}
 \caption{The density distribution of the massive WD mass ($M_{\rm WD1}$), the less massive WD mass ($M_{\rm WD2}$), the total mass ($M_{\rm WD1}+M_{\rm WD2}$) and the mass ratio ($M_{\rm WD2}/M_{\rm WD1}$) of the double WDs as a function of time. Here, every time-bin has been normalized to 1.}
  \end{center}
\end{figure*}

The distribution of the masses of WD mergers for producing SNe Ia varies with time. Fig.\,7 shows the evolution of the massive WD mass $M_{\rm WD1}$ (the top left panel), the less-massive WD mass $M_{\rm WD2}$ (the top right panel), the total mass $M_{\rm WD1}+M_{\rm WD2}$ (the bottom left panel) and the mass ratio $M_{\rm WD2}/M_{\rm WD1}$ (the bottom right panel) as a function of time. All of these DD systems potentially merge to form SNe Ia. The total mass of WD mergers are in the range of $0.8-1.0\,\rm M_{\odot}$, and their mass ratio are mainly concentrated between 0.6 and 0.7. Note that Yungelson \& Kuranov (2017) also presented a similar distribution in their Fig.\,5. The mainly difference between these two works are as follows: (1) Yungelson \& Kuranov (2017) have not considered the WD$+$He subgiant channel for the production of SNe Ia. (2) Yungelson \& Kuranov (2017)  assumed that the merging of CO WDs more massive than $0.47\,\rm M_{\odot}$ with a He or hybrid HeCO WDs more massive than $0.37\,\rm M_{\odot}$ can produce SN Ia. This merger scenario is not included in our CE ejection channels. Obviously the total mass of these mergers could be less than the Chandrasekhar limit for producing SNe Ia (see also Liu et al. 2017).

\section{Discussion and Summary}\label{Discussion and Summary}
\begin{figure}
\begin{center}
\epsfig{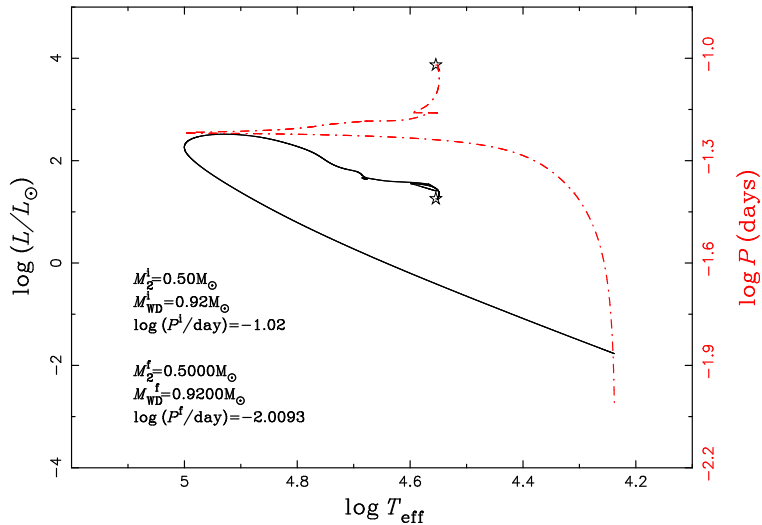}
 \caption{Similar to the right panel of Fig.\,1, but for the particular evolution of a CO WD$+$sdB system KPD 1930$+$2752.}
  \end{center}
\end{figure}
In the observations, there are many candidates for SN Ia progenitors, in which KPD 1930$+$2752 is a massive WD$+$sdB star system and may evolve to form an SN Ia via the DD model (Maxted et al. 2000; Geier et al. 2007). Downes (1986) firstly identified KPD 1930$+$2752 as a sdB star. Bill\'eres et al. (2000) found that the sdB star has a massive companion, and predicted that the orbital period of KPD 1930$+$2752 is about $2.83\,\rm h$. Geier et al. (2007) suggested that the mass of the sdB star in KPD 1930$+$2752 is in the range of $0.45-0.52\,\rm M_{\odot}$, and the total mass of the binary ranges from $1.36$ to $1.48\,\rm M_{\odot}$, which is likely to exceed the Chandrasekhar limit.
In Fig\,8, we present the evolution of the sdB star in KPD 1930$+$2752. Here, the solid curve shows the luminosity evolution of the sdB star as a function of effective temperature, and the dash-dotted curve represents the evolution of the binary orbital period. From our calculations, we found that KPD 1930$+$2752 will not experience mass transfer process until the formation of a DD system. It takes $\sim$$214\,\rm Myr$ for KPD 1930$+$2752 to evolve to double WDs, and the DD system will merge in about $4\,\rm Myr$.

In addition, some He stars with masses lower than $0.6\,\rm M_{\odot}$ will not experience mass transfer process and directly evolve to WDs. The He star with low mass ($\sim$$0.45$$-$$0.6\,\rm M_{\odot}$, see Dan et al. 2012, 2014) will evolve to a HeCO hybrid WD that have a CO core surrounded by a He mantle (e.g. Iben \& Tutukov 1985; Tutukov \& Yungelson 1996; Han, Tout \& Eggleton 2000). Thus, alternatively the outcome of the sdB star in KPD 1930$+$2752 might be a HeCO hybrid WD. The merging of a HeCO hybrid WD$+$CO WD may also lead to the production of SNe Ia via the double detonation model, in which the surface explosion of the He-rich shell drives a shock compression onto the CO WD, leading to another explosion of the whole CO WD (e.g., Guillochon et al. 2010; Dan et al. 2012, 2014, 2015; Pakmor et al. 2013; Papish et al. 2015; Liu et al. 2017).

\begin{figure*}
\begin{center}
\epsfig{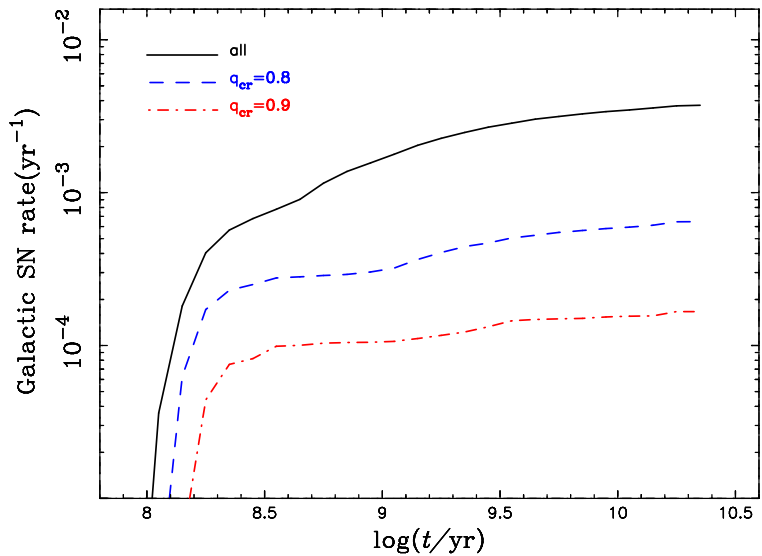}
 \caption{The influence of different critical mass ratio of WD mergers on the birthrates (the left panel) and delay time distributions (the right panel) of SNe Ia. Here, the $\alpha_{\rm CE}\lambda$ is set to be 1.0.}
  \end{center}
\end{figure*}

It has been supposed that the DD model can roughly reproduce the birthrates and delay time distributions of SNe Ia in observations (see also Ruiter et al. 2009; Yungelson \& Kuranov 2017).
However, the outcome of these merging of double CO WDs is still under debate. Some theoretical studies argued that the double WD merger may lead to an accretion-induced collapse and finally forms a neutron star as the mass transfer rate is relatively high (e.g. Saio \& Nomoto 1985, 1998; Kawai, Saio \& Nomoto 1987; Timmes, Woosley \& Taam 1994; Shen et al. 2012; Schwab, Quataert \& Kasen 2016). Recently, it has been proposed that the accretion-induced collapse may be avoided when the merger of double WDs is violent (Pakmor et al. 2010, 2011, 2012). The mass ratio of WD mergers have a great influence on the outcome of the merging of double WDs (e.g. Sato et al. 2016); an SN Ia would be produced when the mass ratio of WD mergers larger than a critical value based on the violent merger scenario ($\sim$0.8; see Pakmor et al. 2011; Liu et al. 2016).
In Fig.\,9, we present the Galactic birthrates of SNe Ia as a function of time based on different critical mass ratios of WD mergers ($q_{\rm cr}$), in which the $q_{\rm cr}$ is set to be 0.8, 0.9 or unlimited. The critical mass ratio $q_{\rm cr}=0.8$ is consistent with previous assumptions (e.g. Pakmor et al. 2011; Liu et al. 2016), which is, however, somewhat optimistic. Thus, we also present the case with $q_{\rm cr}=0.9$ for comparison. Here, we show the results from all channels and set $\alpha_{\rm CE}\lambda=1$ (see Table\,1). The birthrate of SNe Ia is $\sim$$0.65\times10^{\rm -3}\,\rm yr^{\rm -1}$ for the case with $q_{\rm cr}=0.8$, and $\sim$$0.17\times10^{\rm -3}\,\rm yr^{\rm -1}$ for the case with $q_{\rm cr}=0.9$, corresponding to about 16\% and 5\% of all SNe Ia, respectively.

In the present work, we evolved a large number of WD$+$He star systems until the merging time of double WDs, and thus obtained a dense and large grid for producing SNe Ia based on the DD model. Using these binary evolution results, we carried out a series of Monte Carlo BPS simulations and found that the DTDs of SNe Ia fit better with that of observations after considering the WD$+$He subgiant channel, especially for old epoches and early epoches that cannot be reproduced by previous studies.
Our standard model shows that the Galactic SN Ia birthrate from the WD$+$He subgiant channel is about $2.4\times10^{\rm -3}\,\rm yr^{\rm -1}$, and the total birthrate of SNe Ia from all channels of the DD model is about $3.7\times10^{\rm -3}\,\rm yr^{\rm -1}$ that is consistent with the observed results. The WD$+$He subgiant channel is one of the most important channel for the formation of massive DD systems, which may corresponding to double DB/DO WDs. In order to provide constraints on the DD model of SNe Ia, large samples of observed WD$+$He star systems and massive double WDs are expected, and further numerical researches on the double WD mergers are needed.

\section*{Acknowledgments}
The present work is supported by the National Basic Research Program of China (973 programme, 2014CB845700), the Natural Science Foundation of China (Nos. 11673059, 11521303, 11573016 and 11390374),
Chinese Academy of Sciences (Nos. KJZD-EW-M06-01 and QYZDB-SSW-SYS001) and Yunnan Province (Nos. 2013HA005, 2013HB097 and 2017HC018).

\label{lastpage}

\begin{thebibliography}{}\label{thebibliography}
\bibitem[Ashok \& Banerjee (2003)]{Ash03}        Ashok N. M., Banerjee D. P. K., 2003, A\&A, 409, 1007
\bibitem[Barstow et al. (1995)]{Bar95}           Barstow M. A., Jordan S., O¡¯Donoghue D. et al., 1995, MNRAS, 277, 971
\bibitem[Bill\'eres et al. (2000)]{Bil00}        Bill\'eres M., Fontaine G., Brassard P. et al., 2000, ApJ, 530, 441
\bibitem[Cappellaro \& Turatto (1997)]{Cap97}    Cappellaro E., Turatto M., 1997, in Thermonuclear Supernovae, ed. P. Ruiz-Lapuente, R. Cannal, \& J. Isern (Dordrecht: Kluwer), 77
\bibitem[Chen \& Li (2007)]{CL07}                Chen W. C., Li X. D., 2007, ApJL, 658, 51
\bibitem[Chen et al. (2012)]{Che12}              Chen X., Jeffery C. S., Zhang X., Han Z., 2012, ApJ, 755, L9
\bibitem[Dan et al. (2015)]{Dan15}               Dan M., Guillochon J., Br\"uggen M., Ramirez-Ruiz E., Rosswog S., 2015, MNRAS, 454, 4411
\bibitem[Dan et al. (2012)]{Dan12}               Dan M., Rosswog S., Guillochon J., Ramirez-Ruiz E., 2012, MNRAS, 422, 2417
\bibitem[Dan et al. (2014)]{Dan14}               Dan M., Rosswog S., Br\"ugen M., Podsiadlowski Ph., 2014, MNRAS, 438, 14
\bibitem[Downes (1986)]{Dow86}                   Downes R. A. 1986, ApJS, 61, 569
\bibitem[Eggleton (1973)]{egg73}                 Eggleton P. P., 1973, MNRAS, 163, 279
\bibitem[Eggleton, Fitchett \& Tout (1989)]{Egg89} Eggleton P. P., Fitchett M. J., Tout C. A., 1989, ApJ, 347, 998
\bibitem[Eggleton et al. (2002)]{egg02}          Eggleton P. P., Kiseleva-Eggleton L., 2002, ApJ, 575, 461
\bibitem[Geier et al. (2013)]{Gei13}             Geier S., Marsh T. R., Wang B. et al., 2013, A\&A, 554, A54
\bibitem[Geier et al. (2007)]{Gei07}             Geier S., Nesslinger S., Heber U., Przybilla N., Napiwotzki R., Kudritzki R. P., 2007, A\&A, 464, 299
\bibitem[Graham et al. (2015)]{Gra15}            Graham M. L. et al., 2015, MNRAS, 454, 1948
\bibitem[Graur \& Maoz (2013)]{Gra13}            Graur O., Maoz D., 2013, MNRAS, 430, 1746
\bibitem[Guillochon et al. (2010)]{Gui10}        Guillochon J., Dan M., Ramirez-Ruiz E., Rosswog S., 2010, ApJL, 709, L64
\bibitem[Hachisu, Kato \& Nomotobinary evolution calculations (1996)]{hac96}  Hachisu I., Kato M., Nomoto K., 1996, ApJ, 470, L97
\bibitem[Han, Podsiadlowski \& Eggleton (1994)]{HAN94}  Han Z., Podsiadlowski Ph., Eggleton P. P., 1994, MNRAS, 270, 121
\bibitem[Han, Podsiadlowski \& Eggleton (1995)]{HAN95}  Han Z., Podsiadlowski Ph., Eggleton P. P., 1995, MNRAS, 272, 800
\bibitem[Han (1998)]{han98}                      Han Z., 1998, MNRAS, 296, 1019
\bibitem[Han, Tout \& Eggleton (2000)]{han00}    Han Z., Tout C. A., Eggleton P. P., 2000, MNRAS, 319, 215
\bibitem[Han \& Podsiadlowski (2004)]{han04}     Han Z., Podsiadlowski Ph., 2004, MNRAS, 350, 1301
\bibitem[Hichen et al. (2007)]{Hic07}            Hicken M. et al., 2007, ApJ, 669, L17
\bibitem[H\"oflich et al. (2013)]{Hof13}         H\"oflich P., Dragulin P., Mitchell J. et al., 2013, Frontiers of Physics, 8, 144
\bibitem[H\"oflich \& Khokhlov (1996)]{Hof96}    H\"oflich P., Khokhlov A., 1996, ApJ, 457, 500
\bibitem[Hillebrandt \& Niemeyer (2000)]{Hil00}  Hillebrandt W., Niemeyer J. C., 2000, ARA\&A, 38, 191
\bibitem[Horesh et al. (2012)]{Hor12}            Horesh A. et al., 2012, ApJ, 746, 21
\bibitem[Howell (2011)]{how11}                   Howell D. A., 2011, Nature Commun, 2, 350
\bibitem[Howell et al. (2006)]{How06}            Howell D. A. et al., 2006, Nature, 443, 308
\bibitem[Hoyel \& Fowler (1960)]{HF60}           Hoyel F., Fowler W. A., 1960, ApJ, 132, 565
\bibitem[Hurley, Tout \& Pols (2002)]{Hur02}     Hurley J. R., Tout C. A., Pols O. R., 2002, MNRAS, 329, 897
\bibitem[Iben \& Tutukov (1984)]{IT84}           Iben I., Tutukov A. V., 1984, ApJS, 54, 335
\bibitem[Iben \& Tutukov (1985)]{IT85}           Iben I., Tutukov A. V., 1985, ApJS, 58, 661
\bibitem[Iben \& Tutukov (1989)]{it89}           Iben I., Tutukov A. V., 1989, ApJ, 342, 430
\bibitem[Ivanova et al. (2013)]{iva13}           Ivanova N., Justham S., Avendano Nandez J. L., Lombardi J. C., 2013, Science, 339, 433
\bibitem[Kato \& Hachisu (2004)]{kat04}          Kato M., Hachisu I., 2004, ApJ, 613, L129
\bibitem[Kato et al. (2008)]{kh08}               Kato M., Hachisu I., Kiyota S., Saio H., 2008, ApJ, 684, 1366
\bibitem[Kawai, Saio \& Nomoto (1987)]{Kaw87}    Kawai Y., Saio H., Nomoto K., 1987, ApJ, 315, 229
\bibitem[Kawka et al. (2017)]{Kaw17}             Kawka A., Briggs G. P., Vennes S. et al., 2017, MNRAS, 466, 1127
\bibitem[K\"ulebi et al. (2010)]{Kul10}          K\"ulebi B., Jordan S., Nelan E., Bastian U., Altmann M., 2010, A\&A, 524, A36
\bibitem[Landau \& Lifshitz (1971)]{LL71}        Landau L. D., Lifshitz E. M., 1971, Classical Theory of Fields, Pergamon Press, Oxford
\bibitem[Langer et al. (2000)]{lan00}            Langer N., Deutschmann A., Wellstein S., H\"{o}flich P., 2000, A\&A, 362, 1046
\bibitem[Li \& van den Heuvel (1997)]{li97}      Li X. D., van den Heuvel E. P. J., 1997, A\&A, 322, L9
\bibitem[Liu et al. (2015)]{Liu15}               Liu D., Zhou W., Wu C., Wang B., 2015, RAA, 15, 1813
\bibitem[Liu et al. (2016)]{Liu16}               Liu D., Wang B., Podsiadlowski Ph., Han Z., 2016, MNRAS, 461, 3653L
\bibitem[Liu et al. (2017)]{Liu17}               Liu D., Wang, B., Wu, C., Han Z., 2017, A\&A, arXiv:1707.07387
\bibitem[Liu et al. (2012)]{liuj12}              Liu J. F., Di Stefano  R., Wang T., Moe  M. 2012, ApJ, 749, 141
\bibitem[Liu et al. (2010)]{Liu10}               Liu W. M., Chen W. C., Wang B., Han Z., 2010, A\&A, 523, 3
\bibitem[Livne (1990)]{Liv90}                    Livne E., 1990, ApJ, 354, L53
\bibitem[L\"u et al. (2009)]{Lv09}               L\"u G., Zhu C., Wang Z., Wang N., 2009, MNRAS, 396, 1086
\bibitem[Maoz, Keren \& Avishay (2010)]{mao10}   Maoz D., Keren S., Avishay G. Y., 2010, ApJ, 722, 1879
\bibitem[Maoz, Mannucci\& Nelemans (2014)]{mao14} Maoz D., Mannucci F., Nelemans G., 2014, ARA\&A, 52, 107
\bibitem[Maoz, Mannucci \& Timothy (2012)]{mao12} Maoz D., Mannucci F., Timothy D. B., 2012, MNRAS, 426, 3282
\bibitem[Maxted et al. (2000)]{Max00}            Maxted P. F. L., Marsh T. R., North R. C., 2000, MNRAS, 317, L41
\bibitem[Meng, Gao \& Han (2015)]{men15}         Meng X., Gao Y., Han Z., 2015, IJMPD, 24, 14, 1530029
\bibitem[Meng \& Yang (2012)]{men12}             Meng X., Yang W., 2012, A\&A, 543, A137
\bibitem[Mereghetti et al. (2009)]{mer09}        Mereghetti S., Tiengo A., Esposito P. et al., 2009, Science, 325, 1222
\bibitem[Mereghetti et al. (2011)]{Mer11}        Mereghetti S., Palombara La N., Tiengo A. et al., 2011, ApJ, 737, 51
\bibitem[Miller \& Scalo (1979)]{mil79}          Miller G. E., Scalo J. M., 1979, ApJS, 41, 513
\bibitem[Nelemans et al. (2001)]{Nel01}          Nelemans G., Yungelson L. R., Portegies Zwart S. F., Verbunt F., 2001, A\&A, 365, 491
\bibitem[Neunteufel, Yoon \& Langer (2016)]{Neu16} 	Neunteufel P., Yoon S.-C., Langer N., 2016, A\&A, 589, 43
\bibitem[Nomoto (1982)]{nom82}                   Nomoto K., 1982, ApJ, 253, 798
\bibitem[Nomoto et al. (1984)]{Nom84}            Nomoto K., Thielemann F. K., Yokoi K., 1984, ApJ, 286, 644
\bibitem[Pakmor et al. (2010)]{Pak10}            Pakmor R., Kromer M., R\"{o}pke F. K., Sim S. A., Ruiter A. J., Hillebrandt W., 2010, Nature, 463, 61
\bibitem[Pakmor et al. (2011)]{Pak11}            Pakmor R., Hachinger S., R\"{o}pke F. K., Hillebrandt W., 2011, A\&A, 528, A117
\bibitem[Pakmor et al.(2012)]{Pak12}             Pakmor R., Kromer M., Taubenberger S. et al., 2012, ApJL, 747, L10
\bibitem[Pakmor et al. (2013)]{Pak13}            Pakmor R., Kromer M., Taubenberger S., Springel V., 2013, ApJL, 770, L8
\bibitem[Papish et al. (2015)]{Pap15}            Papish O., Soker N., Garc\'ia-Berro E., Aznar-Siguan G., 2015, MNRAS, 449, 942
\bibitem[Podsiadlowski et al. (2008)]{Pod08}     Podsiadlowski Ph., Mazzali P., Lesaffre P., Han Z., F\"orster F., 2008, New Astro. Rev., 52, 381
\bibitem[Pols et al. (1998)]{pol98}              Pols O. R., Schr\"{o}der K. P., Hurly J. R., Tout C. A., Eggleton P. P., 1998, MNRAS, 298, 525
\bibitem[Pols et al. (1995)]{pol95}              Pols O. R., Tout C. A., Eggleton P. P., Han Z., 1995, MNRAS, 274, 964
\bibitem[R\"{o}pke \& Hillebrandt (2005)]{rop05} R\"{o}pke F. K., Hillebrandt W., 2005, A\&A, 431, 635
\bibitem[Ruiter et al. (2009)]{Rui09}            Ruiter A. J., Belczynski K., Fryer C., 2009, ApJ, 699, 2026
\bibitem[Ruiter et al. (2013)]{Rui13}            Ruiter A. J. et al., 2013, MNRAS, 429, 1425
\bibitem[Saio \& Nomoto (1985)]{Sai85}           Saio H., Nomoto K., 1985, A\&A, 150, 21
\bibitem[Saio \& Nomoto (1998)]{Sai98}           Saio H., Nomoto K., 1998, ApJ, 500, 388
\bibitem[Santander-Garc\'ia et al. (2015)]{san15} Santander-Garc\'ia M. et al., 2015, nature, 519, 63
\bibitem[Sato et al. (2016)]{Sat16}              Sato Y. et al., 2016, ApJ, 821, 67
\bibitem[Scalzo et al. (2010)]{Sca10}            Scalzo R. A. et al., 2010, ApJ, 713, 1073
\bibitem[Schwab, Quataert \& Kasen (2016)]{Sch16} Schwab J., Quataert E., Kasen D., 2016, MNRAS, 463, 3461
\bibitem[Shen et al. (2012)]{She12}              Shen K. J., Bildsten L., Kasen D., Quataert E., 2012, ApJ, 748, 35
\bibitem[Timmes et al. (1994)]{Tim94}            Timmes F. X., Woosley S. E., Taam R. E., 1994, ApJ, 420, 348T
\bibitem[Toonen et al. (2012)]{Too12}            Toonen S., Nelemans G., Portegies Z. S., 2012, A\&A, 546, A70
\bibitem[Totani et al. (2008)]{tot08}            Totani T., Morokuma T., Oda T., Doi M., Yasuda N., 2008, PASJ, 60, 1327
\bibitem[Tout et al. (2008)]{Tou08}              Tout C. A., Wickramasinghe D. T., Liebert J., Ferrario L., Pringle J. E., 2008, MNRAS, 387, 897
\bibitem[Tutukov \& Yungelson (1996)]{Tut96}     Tutukov A., Yungelson L., 1996, MNRAS, 280, 1035
\bibitem[Wang et al. (2009b)]{Wan09b}            Wang B., Chen X., Meng X., Han Z., 2009b, ApJ, 701, 1540
\bibitem[Wang \& Han (2010)]{wan10}              Wang B., Han Z., 2010, RAA, 681, 688
\bibitem[Wang \& Han (2012)]{wh12}               Wang B., Han Z., 2012, New Astron. Rev., 56, 122
\bibitem[Wang, Justham \& Han (2013)]{wan13}     Wang B., Justham S., Han Z., 2013, A\&A, 559, A94
\bibitem[Wang et al. (2009a)]{Wan09a}            Wang B., Meng X., Chen X., Han Z., 2009a, MNRAS, 395, 847
\bibitem[Wang, Podsiadlowski \& Han (2017)]{Wan17} Wang B., Podsiadlowski Ph., Han Z., 2017, MNRAS, in press (arXiv:1708.07067)
\bibitem[Webbink (1984)]{web84}                  Webbink R. F., 1984, ApJ, 277, 355
\bibitem[Whelan \& Iben (1973)]{whe73}           Whelan J., Iben I., 1973, ApJ, 186, 1007
\bibitem[Willems \& Kolb (2004)]{Will04}         Willems B., Kolb U., 2004, A\&A, 419, 1057
\bibitem[Woosley, Taam \& Weaver (1986)]{Woo86}  Woosley S. E., Taam R. E., Weaver T. A., 1986, ApJ, 301, 601
\bibitem[Woudt et al. (2009)]{Wou09}             Woudt, P. A., Steeghs D., Karovska M. et al., 2009, ApJ, 706, 738
\bibitem[Yoon \& Langer (2003)]{yoo03}           Yoon S. C., Langer N., 2003, A\&A, 412, L53
\bibitem[Yoon Podsiadlowski \& Rosswog (2007)]{Yoon07} Yoon S.-C., Podsiadlowski Ph., Rosswog S., 2007, MNRAS, 390,933
\bibitem[Yungelson \& Kuranov (2017)]{YK17}      Yungelson L. R., Kuranov A. G., 2017, MNRAS, 464, 1607
\bibitem[Yungelson \& Livio (1998)]{YL98}        Yungelson L. R., Livio M., 1998, ApJ, 497, 168
\end{thebibliography}
\end{document}